\documentclass[twocolumn,twoside,slac_two]{revtex4}
\usepackage{amsmath}
\usepackage{amsfonts}
\usepackage{amssymb}
\usepackage[dvips]{graphicx}
\usepackage[dvips]{graphicx,color}
\usepackage{epsfig}
\usepackage{fancyhdr}
\newcommand{\DZ}{D^0}
\newcommand{\DZB}{\overline{D^0}}

\newcommand{\etal}{{\it et al.}}
\newcommand{\kspipi}{K_{S}\pi^{+}\pi^{-}}
\newcommand{\klpipi}{K_{L}\pi^{+}\pi^{-}}

\begin{document}

\title{Impact on \boldmath $\gamma/\phi_3$ from CLEO-c 
Using $CP$-Tagged $D\rightarrow K_{S,L}\pi^{+}\pi^{-}$ Decays \unboldmath}

%
 
\author{Eric White$^1$, Qing He$^2$, for the CLEO Collaboration}
\affiliation{$^1$University of Illinois, Urbana-Champaign, IL 61801 USA,
$^2$University of Rochester, Rochester, NY 14627 USA}

\begin{abstract}
Precision determination of the CKM angle $\gamma/\phi_3$ depends upon
constraints on charm mixing amplitudes, measurements of doubly-Cabibbo
suppressed amplitudes and relative phases, and studies of charm Dalitz
plots tagged by flavor or $CP$ eigenstates. 
In this note we describe the technique used at CLEO-c to constrain the $K_{S,L}\pi^+\pi^-$ model
uncertainty, and its impact on $\gamma/\phi_3$ measurements at 
$B$-factories presented at the Charm~2007 Workshop.
\end{abstract}
\maketitle

\thispagestyle{fancy}


\section{Introduction}

Measurement of the CKM angle $\gamma/\phi_3$ is challenging. 
Several methods have been proposed using $B^\mp \to D K^\mp$ decays;
 1) the
Gronau-London-Wyler (GLW) method~\cite{glw} where the $D$ decays to $CP$
eigenstates
2) the Atwood-Dunietz-Soni (ADS) method~\cite{ads} where the $D$ decays to
flavor eigenstates
and 3) the Dalitz plot method~\cite{bondar,ggsz} 
where the $D$ decays to a three-body final state. This latter method has been used 
recently by CLEO to
measure the $K^{*}K$ strong phase via the three-body decay
$\DZ \rightarrow K^+K^-\pi^0$ \cite{paras}. 
Uncertainties due to charm contribute to each of these methods.
The CLEO-c physics program includes a variety of charm
measurements
that will improve the determination of $\gamma/\phi_3$ from the
$B$-factory experiments, BaBar and Belle.
The pertinent components of this program are improved constraints on charm
mixing amplitudes - important for GLW, measurement of the relative strong
phase $\delta$ between
$\DZ$ and $\DZB$ decay to $K^+\pi^-$ - important for ADS, and studies of
charm Dalitz plots tagged by hadronic flavor or $CP$ eigenstates. 
The total number of charm mesons accumulated at CLEO-c will be much
smaller
than the samples already accumulated by the $B$-factories. However,
quantum correlations in the $D\bar D$ system from $\psi(3770)$ provides
a unique laboratory in which to study charm.

The decay with the largest branching fraction relevant to the determination of $\gamma/\phi_3$
$\DZ \rightarrow \kspipi$. Recently
Babar~\cite{babar} and Belle~\cite{belle} have
reported $\gamma = (92\pm 41 \pm 11 \pm 12)^\circ$ and
$\phi_3 = (53^{+15}_{-18} \pm 3 \pm 9)^\circ  $, respectively, where the
third error is the
systematic error due to modeling of the Dalitz plot.

Both $\DZ$ and $\DZB$ populate the Dalitz plots $\kspipi$,
(as well as $\pi^+\pi^-\pi^0$, $K^+K^-\pi^0$ and $K^0_S K^\pm\pi^\mp$) and
so can be
used in the determination of $\gamma/\phi_3$ which exploit the
interference between
$b \to c \bar u s$ ($B^- \to \DZ K^-$) and $b \to u \bar c s$
($B^- \to \DZB K^-)$ where the former process is real and the latter
is proportional to $\sim e^{-i\gamma}$~\cite{bigi}. Studying $CP$ tagged Dalitz plots
allows a model independent determination of the relative $\DZ$ and $\DZB$
phase across the Dalitz plot. We describe this technique in the following
sections.

\section{Determining \boldmath $\gamma/\phi_3$ From $B$ \unboldmath Decays}

Our analysis follows the work outlined in \cite{bondar}, \cite{ggsz}, and \cite{bondar2}.
We consider the decay process $B^{\pm}\rightarrow DK^{\pm}$, followed by the three-body
decay $D\rightarrow \kspipi$. Assuming no $CP$ violation, 
we define the decay amplitudes for the $\DZ$ and $\DZB$ to be
\begin{equation}
\begin{array}{rcl}
\mathcal{A}(\DZ \rightarrow K_S\pi^+\pi^-; x,y) &\equiv& f_D(x,y) \\
\mathcal{A}(\DZB \rightarrow K_S\pi^-\pi^+; x,y) &\equiv& f_D(y,x).
\\ & & \\
\end{array}\label{eq-def}
\end{equation}
Sensitivity to the angle $\phi_3$ comes from the interference of the neutral 
$D$ mesons from $B^{\pm} \rightarrow DK^{\pm}$. 
Since the $D$ meson is in a linear
combination of flavor states, the amplitude for a $\DZ \rightarrow 
\kspipi$ event originating from a $B$ decay is then
\begin{equation}
\mathcal{A}(B^- \!\!\rightarrow\! (K_S\pi^+\pi^-)_DK^-) \propto f_D + r_B e^{i\theta_-}f_{\overline D},
\end{equation}
up to an overall normalization. The angle $\theta$ is
defined as $\theta_{\pm} \equiv \delta_B \pm \phi_3$. Here $\delta_B$ is the strong phase
difference between color-suppressed and favored amplitudes, 
whilst $r_B$ is the ratio between the color-suppressed to favored amplitudes. Theoretical estimates place $r_B$
between 0.1-0.2 \cite{gronau}. This has been confirmed by BaBar
($r_B$ = 0.12 $\pm$ 0.08 $\pm$ 0.03(syst) $\pm$ 0.04(model), \cite{babar}) and Belle
($r_B$ = 0.16 $\pm$ 0.05 $\pm$ 0.01(syst) $\pm$ 0.05(model), \cite{belle}).

\begin{figure}[htbp]
\begin{center}
\epsfig{file=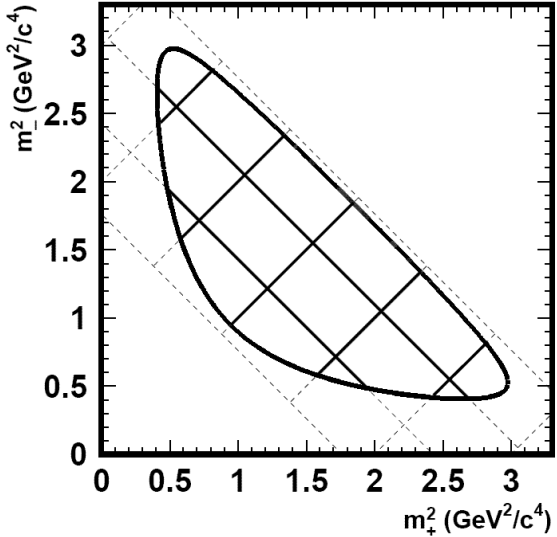, width=75mm}
\end{center}
\caption{Binning of the $\DZ\rightarrow\kspipi$ Dalitz plot. }
\label{fig1}
\end{figure}

The $\DZ \rightarrow \kspipi$ Dalitz plot is divided into $2\mathcal{N}$ bins,
 symmetric under exchange of $x$ and $y$. 
The bins are indexed from $-i$ to $i$, excluding zero, as in shown in Fig~\ref{fig1}.
The coordinate exchange $x \leftrightarrow y$ thus corresponds to the exchange of bins $i \leftrightarrow -i$.
For simplicity we ignore the effects of efficiency and background in the Dalitz plot.
The number of events in the $i$-th bin of the $\kspipi$ Dalitz plot from a $D$ decay is then expressed as
\begin{equation}\label{Ksubi}
K_i = A_D \int_{\mathcal{D}_i}\left|f_D(x,y)\right|^2 dx\ dy = A_DF_i.
\end{equation}

The interference between the $\DZ$ and $\DZB$ amplitudes is parametrized by the two quantities
\begin{equation}
c_i \equiv \frac{1}{\sqrt{F_iF_{-i}}} \int_{\mathcal{D}_i}\text{Re}\left[f_D(x,y)f_D^{*}(y,x)\right] dx\ dy
\end{equation}
and
\begin{equation}
s_i \equiv \frac{1}{\sqrt{F_iF_{-i}}} \int_{\mathcal{D}_i}\text{Im}\left[f_D(x,y)f_D^{*}(y,x)\right] dx\ dy,
\end{equation}
where the integral is performed over a single bin. 
The number of events in the $i$-th bin of the $\kspipi$ Dalitz plot from a $B$ decay is then
\begin{eqnarray}
N_i &=& K_i + r^2_B K_{-i} \nonumber \\
 & & - 2r_B\sqrt{K_iK_{-i}}\left(c_i\cos{\theta_-} - s_i\sin{\theta_-} \right),
\end{eqnarray}
again up to an overall normalization. 
It is important to note that $c_i$ and $s_i$ depend only on the $D$ decay. These
are the quantities that we measure using CLEO-c data.
 Although in principle they could be left as free parameters in a
 $D \rightarrow \kspipi$ Dalitz plot analysis from $B^{\pm}$ decays, their values
can be more precisely determined from correlated $D_{CP}$ decays produced at CLEO-c.

Thus, we can constrain $\theta_{\pm}$, and in turn $\gamma/\phi_3$, if we know 
$K_{i}$, $c_i$, and $s_i$. 
The $K_{i}$ can be easily determined using flavor-tagged $\DZ \rightarrow \kspipi$ Dalitz plot.
In the next section we show how the $c_i$ can  be obtained using binned,
$CP$-tagged $\DZ \rightarrow \kspipi$ Dalitz plots.



\section{Measuring \boldmath $c_i$ From $CP$-Tagged $D$ \unboldmath Decays}

For $D$ mesons that decay into a $CP$ eigenstate, we write the initial state of the $D$ as
a linear combination of flavor eigenstates
\begin{equation}
  f_{CP\pm}(x,y) = \frac{1}{\sqrt{2}}\left|f_D(x,y) \pm f_D(y,x)\right|.
\end{equation}
In terms of this amplitude the number of events in the $i$-th bin of a $CP$-tagged Dalitz plot is
\begin{equation}
M^{\pm}_{i} = h_{CP\pm}\left(K_i \pm 2c_i\sqrt{K_iK_{-i}} + K_{-i} \right),
\label{Msubi}
\end{equation}
where $h_{CP\pm}$ is a normalization factor.

The expression given above for $M^{\pm}_i$ can be used to measure $c_i$ directly,
 even if only one type of $CP$ tag is reconstructed. Care must be taken to use the
corresponding value of $h_{CP\pm}$ as defined above. However,
if samples of both $CP$ parities are available
we can combine the expressions for $M^+_i$ and $M^-_i$ to get the following equation
\begin{equation}
c_i = \frac{1}{2}\frac{\left(M^-_i - M^+_i\right)}
{\left(M^+_i + M^-_i\right)}
\frac{\left(K_i + K_{-i}\right)}{\sqrt{K_iK_{-i}}},
\label{equationCi}
\end{equation}
We thus have an expression for measuring $c_i$ simply by counting events within the bins
of flavor-tagged and $CP$-tagged Dalitz plots.

At CLEO-c we produce $\DZ\DZB$ pairs from the decay of a $\psi(3770)$ 
in a definite eigenstate of $C = -1$. 
Ignoring both the effects of $CP$ violation, 
the double tag rate for final states $|1\rangle$ and $|2\rangle$ is given by
\begin{equation}
\Gamma(1,2) = |A(1,2)|^2 + \text{(mixing terms)},
\end{equation}
where
\begin{equation}
A(1,2) \equiv \langle 1|\DZ\rangle \langle 2|\DZB\rangle 
     - \langle 1|\DZB\rangle \langle 2|\DZ\rangle
\end{equation}
For the time being we ignore the effects of correlations and mixing in the $K\pi$ tagged Dalitz plot.
This is not expected to make a significant
difference for the $K\pi$ mode, as terms proportional to $r_{K\pi} \simeq 0.06$ and 
$r_{K\pi}^2$ are negligible.

\subsection{Optimized Binning}

Although the quantity $s_i$ can only be measured using a $\kspipi$ vs. $\kspipi$ double Dalitz 
analysis~\cite{bondar2},
 it can still be approximated from a single Dalitz plot if the binning is fine enough. 
If the bins are small enough that the phase difference and the amplitude remains constant across each bin, 
the strong phase parameters become $c_i = \cos{(\delta_D)}$, $s_i = \sin{(\delta_D)}$,
so that the equality $s_i = \sqrt{1 - c_i^2}$ is true.
It has been shown \cite{bondar} that this equality holds for 200
or more bins, which is clearly not feasible for the number of $D_{CP}$ tags produced
at CLEOc.
In order to circumvent this problem, Bondar has proposed an alternate,
model-dependent method for binning the $\kspipi$ Dalitz plot\cite{bondar2}. 
The optimal choice depends on the $D^0\rightarrow\kspipi$ model. 
In this analysis we use the isobar model amplitude obtained from the most recent 
Belle $\phi_3$ Dalitz analysis \cite{belle}.

\begin{figure}[htbp]
\begin{center}
\epsfig{file=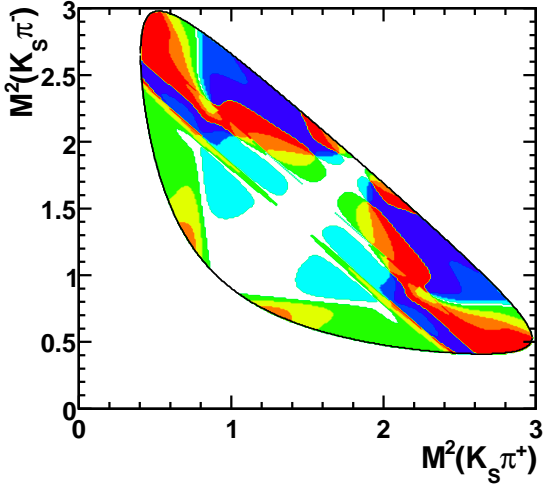, width=75mm}
\end{center}
\caption{Divisions of the $\DZ\rightarrow\kspipi$ Dalitz plot with uniform binning of
$\Delta\delta_D$ strong phase difference with  N = 8. }
\label{fig2}
\end{figure}

From the consideration above it is clear that a good approximation to the optimal binning is the
one obtained from the uniform division of the strong phase difference $\delta_D$. We thus take the
definition of $i$-th bin to be
\begin{equation}
2\pi(i - 1/2)/N \le \delta_D(x,y) < 2\pi(i + 1/2)/N.
\end{equation}
An example of such a binning with
$N = 8$ is shown in Fig.~\ref{fig2}.


\section{Event Selection}

\subsection{Double-Tagged \boldmath $D \rightarrow \kspipi$ \unboldmath Events}

This analysis uses a combination of two-body $CP$ and flavor tags.
Since the neutral $D$ mesons are produced at $\psi^{\prime}(3770)$ threshold
they are correlated in a $C=-1$ state.
If mixing is ignored we can determine whether the parent particle was a $\DZ$ or
$\DZB$, up to DCS contributions. Similarly, if $CP$ violation is ignored, then the $D$ mesons
must be in eigenstates of opposite $CP$~\cite{asnersun}. 

To determine the flavor of the $D$ meson, we tag $\DZ\rightarrow\kspipi$ 
events with the two-body $\DZB\rightarrow K^+\pi^-$ mode.
\footnote{The inclusion of charge-conjugate modes is implied throughout our analysis.} 
We use the two $CP$-even tags $K^+K^-$
and $\pi^+\pi^-$, and the two $CP$-odd tags $K_S\pi^0$ and $K_S\eta$. 

We introduce two quantities that are reconstructed on both sides of a double-tagged decay.
The beam-constrained mass is defined as $M_{bc}~\equiv~\sqrt{E^2_{b} - p_D^2}$, where
$E_b$ is the beam energy and $p_D^2$ is the square of the reconstructed 3-momentum of
the $D$ meson. We require that the beam-constrained mass of the reconstructed candidate is
within 3$\sigma$ of the nominal $D$ mass, which corresponds to a selection criteria of 
$1.8603 \le M_{bc} \le 1.8687$ GeV. 
The other quantity is the energy difference between the beam and the reconstructed $D$,
defined as $\Delta E \equiv E_{beam} - E_{D}$.
We apply a selection criteria of $|\Delta E| \le 30$ MeV to all $\DZ\rightarrow\kspipi$ candidates.

Additional selection criteria are placed on the duaghter particles to ensure basic track quality.
For example, we select pion track momenta between $0.05 \le p \le 2.0$ GeV. 
Both signal and tagging modes containing a $K_S$ are selected to be within $3\sigma$ of the $K_S$ mass, 
which corresponds to $\pm7.5$ MeV from the central $K_S$ mass value of 497.6 MeV.

We only reconstruct $K_S$ particles that decay through the $\pi^+\pi^-$ channel; we do not
attempt to reconstruct $K_S\rightarrow\pi^0\pi^0$. 
Fake $K_S$ candidates can be misreconstructed from combinatoric $\pi^+\pi^-$ pairs.
To suppress these events we apply a selection criteria on the flight significance $f_s \ge 0$ to our $K_S$
candidates. Additionally, we require that the $\pi^0$ mass falls within  3$\sigma$ of its nominal value.

\subsection{Double-Tagged \boldmath $D \rightarrow \klpipi$ \unboldmath Events}

For $\DZ\rightarrow\klpipi$ decays we require the same selection criteria on charged pions and $\pi^0$
candidates as those described for $\DZ\rightarrow\kspipi$ decays. 
However, because of the large flight distance of the $K_L$,
the $\klpipi$ signal is reconstructed using a missing mass technique. 
We require the signal side to have exactly two charged tracks. We also apply $\pi^0$,
$\eta$, and $K_S$ vetoes. Using the measured momentum of the tagged $D$, we compute the
missing momentum and energy on the signal side. 
We require that the missing mass squared satisfies the condition $0.21 \le m^2 \le 0.29$ GeV$^2$.
The background for $\DZ\rightarrow\klpipi$ mode is approximately 5\%.

\subsection{Double-Tagged \boldmath $K_L\pi^0$ vs. $\kspipi$ \unboldmath}

We can increase our statistics by reconstructing $\DZ\rightarrow\kspipi$ events tagged 
with the $CP$-even mode $K_L\pi^0$.
We require zero tracks and exactly one $\pi^0$ candidate on the tag side. 
We veto events containing $\eta$ candidates, and impose similar criteria on the $K_L$ missing mass
as described above.

The final yields for all tag modes are summarized in Table~\ref{table-yield}
\begin{table}[htbp]
\begin{center}
\caption{Yields for CP-tagged $\kspipi$ and $\klpipi$ in 398 pb$^{-1}$ data, by tag mode.}
\begin{tabular}{|l|c|c|}
\hline \textbf{Tag Mode} & \textbf{$\kspipi$} & \textbf{$\klpipi$} \\
\hline $K^+K^-$ & 61 & 194  \\
\hline $\pi^+\pi^-$ & 33 & 90  \\
\hline $K_S\pi^0$ & 108 & 263  \\
\hline $K_S\eta$ & 29 & 21  \\
\hline $K_L\pi^0$ & 190 & -  \\
\hline
\end{tabular}
\label{table-yield}
\end{center}
\end{table}


\section{Combining $\kspipi$ and $\klpipi$}

The tagged $\klpipi$ Dalitz plots are included to increase the statistical accurancy of this analysis.
However, if we naively combine the Dalitz plots with $K_S$ and $K_L$ we will find our measurement
of $c_i$ to be biased. We must first account for the phenomenological differences between the
$\kspipi$ and $\klpipi$ models.

Since the $K_S$ and $K_L$ mesons are of opposite $CP$, the doubly-Cabibbo suppressed amplitudes
in each Dalitz plot will contribute with opposite signs. We can see this by inspecting the $\DZ$ 
decay amplitude for each each Dalitz plot
\begin{eqnarray}
\mathcal A(K_S\pi\pi) &=& \frac{1}{\sqrt{2}}\left[\mathcal A(K^0\pi\pi) + \mathcal A(\overline{K^0}\pi\pi) \right]  \nonumber \\
\mathcal A(K_L\pi\pi) &=& \frac{1}{\sqrt{2}}\left[\mathcal A(K^0\pi\pi) - \mathcal A(\overline{K^0}\pi\pi) \right] \\
\end{eqnarray}

The effect of this relative minus sign is to introduce a 180$^\circ$ phase for all DCS $K^*$ resonances in the
$\klpipi$ model.
We can use $U$-spin symmetry to relate the amplitudes for resonances of definite $CP$ eigenvalue.
We find that these states aquire a factor of $r_K e^{i\delta_K} \simeq -\tan{\theta_C}$, 
where $\theta_C$ is the Cabibbo angle. 

In our study we mulitply all DCS amplitudes in the $\klpipi$ model by -1.
From this ``base'' model we fix $r_K = 0.06$ for each $CP$ eigenstate, then vary the phase $\delta_K$ 
between 0 and $2\pi$. For each bin we then find the largest resulting deviation in $c_i$, and report this value
as the systematic uncertainty in the $\klpipi$ model.
\begin{figure}[htbp]
\begin{center}
\epsfig{file=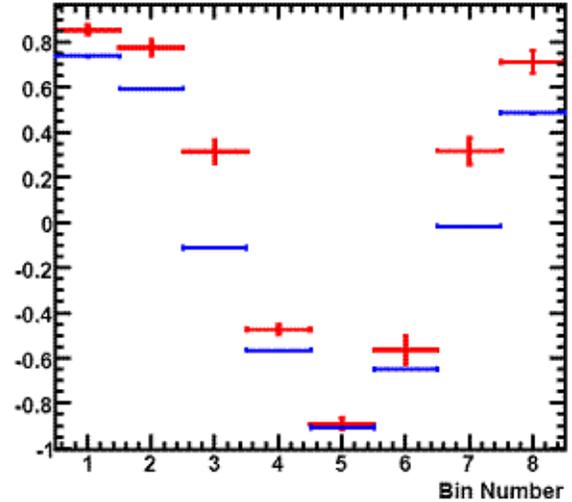, width=75mm}
\end{center}
\caption{Values for $c_i$ numerically determined from our model.
$\klpipi$ values are in red, $\kspipi$ in blue.
In each bin the values for $c_i$ are systematically larger for $\klpipi$. 
The error bars represent the uncertainties in the $\klpipi$ model parametrization. }
\label{fig3}
\end{figure}

To better understand the difference between the $\kspipi$ and $\klpipi$ models,
we compare the numerically calculated values of $c_i$ in each Dalitz plot.
We find that the value for $c_i$ is systematically larger in each bin for $\klpipi$.
In Fig.~\ref{fig3} we can see that the difference is significantly larger than the systematic 
uncertainty in our $\klpipi$ model.

\section{Results}

\begin{figure}[htbp]
\begin{center}
\epsfig{file=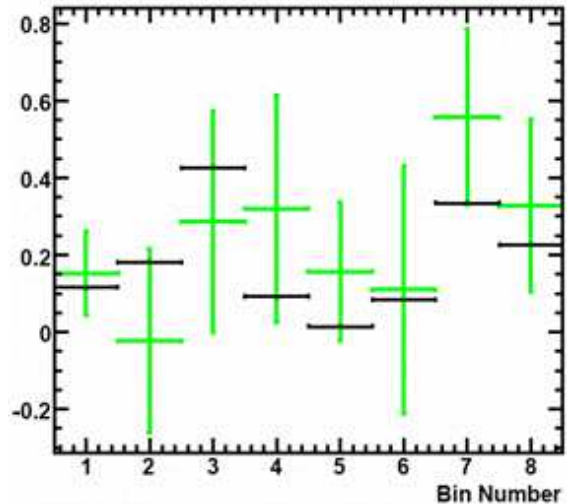, width=75mm}
\end{center}
\caption{The difference in $K_S\pi\pi$ and $K_L\pi\pi$ values of $c_i$, 
numerically determined from our model (black) and measured in 398 pb$^{-1}$ of data (green). 
The green error bars represent the combined statistical and model uncertainty. }
\label{fig4}
\end{figure}

\begin{figure}[htbp]
\begin{center}
\epsfig{file=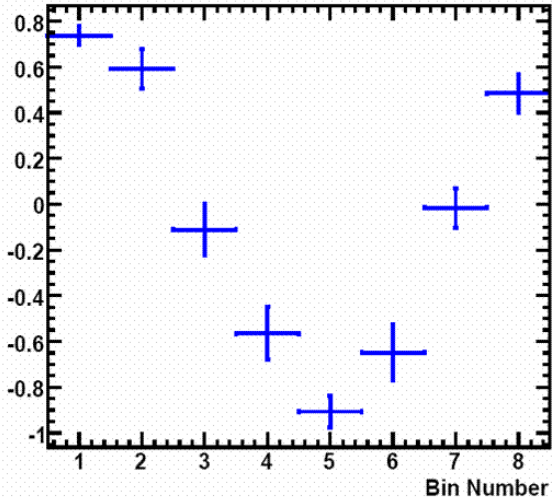, width=75mm}
\end{center}
\caption{The central values of $c_i$ are computed from our model with expected sensitivity from
750 pb$^{-1}$ of data. 
The error bars are determined by scaling the statistical uncertainty obtained from 398 pb$^{-1}$ of data,
then combining the $\klpipi$ model uncertainty.}
\label{fig5}
\end{figure}

We report the difference in $c_i$ between $\kspipi$ and $\klpipi$ as measured in 398 pb$^{-1}$ of data.
In Fig.~\ref{fig4} we compare the $c_i$ differences calculated from our model and measured from data.
The error bars in this figure represent both statistical and model uncertainty combined.
With a reasonable understanding of the $c_i$ between the $\kspipi$ and $\klpipi$ Dalitz plots,
we can estimate the final precision with which we expect to measure $c_i$ once 750 pb$^{-1}$ of data
is available.
The values of $c_i$ from our study are once again plotted in Fig.~\ref{fig5}, but here
the error bars represent the statistical uncertainty obtained from 398 pb$^{-1}$ of data
scaled up to 750 pb$^{-1}$.

We expect good sensitivity to the measurement of $c_i$ with the entire CLEO-c data.
This measurement can reduce the model uncertainty on $\gamma/\phi_3$ to a precision of about $4^\circ$~\cite{bondar}.

\bigskip
\begin{acknowledgments}
We would like to thank Mats Selen from the University of Illinois, our colleagues David Asner and Paras Naik 
from Carleton University, and Ed Thorndike from the University of Rochester for helping 
us prepare for this conference.
Also, we would like to thank the organizers of the Charm 2007 Workshop for providing a stimulating environment
and a well-organized program of talks. 
\end{acknowledgments}

\bigskip 

\end{document}